# ImarisWriter: Open Source Software for Storage of Large Images in Blockwise Multi-Resolution Format

Igor Beati, Eliana Andreica, Peter Majer
Bitplane AG, Zürich, Switzerland

In fluorescence microscopy the size of a single 3D or 4D image may often be larger than the graphics memory available in a typical computer and it may even be larger than the RAM of a typical computer. To visualize such images efficiently it has become standard to rely on blockwise multi-resolution storage formats. These formats allow loading of parts of an image at a preferred resolution which suits well the requirements for visualization. They also allow for building performant image analysis algorithms that process data at a preferred resolution in a block-by-block sequence. Common blockwise and/or multi-resolution formats are BDV [Pietzsch 2015] , KLB [Amat 2015], OME TIFF since version 6.0.0 [Goldberg 2005, OME_TIFF_6.0.0], Zeiss CZI, Imaris IMS, and Arivis SIS. Several of these are based on the low level HDF5 format which is natively blockwise [Folk 2011]. There are benefits and drawbacks to each format but the capabilities of each software that is available to write and read these formats are perhaps more relevant to most users. Here we publish an open source ImarisWriter implementation [ImarisWriter 2020] to make available a high performance writer library for the existing HDF5 based Imaris IMS format [Imaris5 Format]. This library is capable of writing compressed blockwise multi-resolution images with high speed. The library can easily be extended to other formats by adding a new writer implementation. We hope that our library will serve the fluorescence microscopy community well by making high performance writing of images in a useful format very easy.

## 1 The IMS Format For High Performance Visualization And Analysis

The IMS file format [Imaris5 Format] is designed to allow fast visualization and processing of very large images. For this purpose it stores not only the original image data but also lower resolution versions of the original, and it stores them in "small" contiguous 3D blocks[1]. This allows the visualization software to efficiently load only those 3D blocks required for each view. As shown in Figure 1 this can mean that the renderer loads different resolutions in different parts of the view. It can also mean that the renderer skips data outside of the current field of view. Or it can mean that the renderer loads only low resolution data when the zoom level is small. Thereby visualization software packages that perform view dependent rendering (such as Imaris, Arivis, or BigDataViewer) can render images significantly faster than other packages that render the entire high resolution image data and they can render images that exceed a computer's RAM.

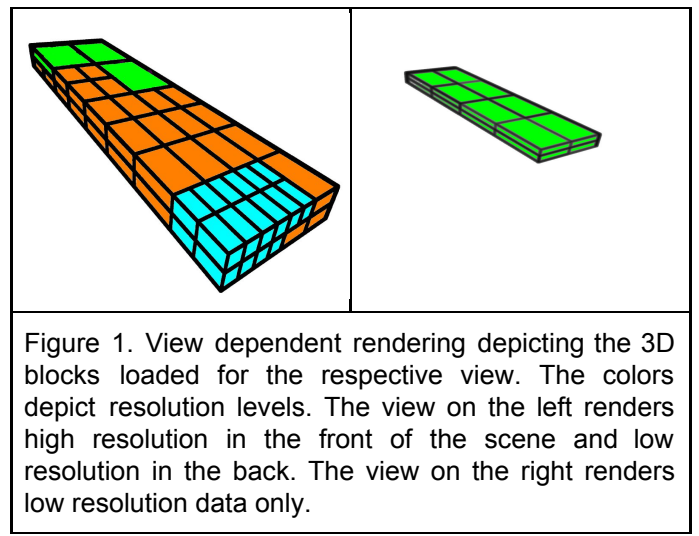

Figure 1. View dependent rendering depicting the 3D blocks loaded for the respective view. The colors depict resolution levels. The view on the left renders high resolution in the front of the scene and low resolution in the back. The view on the right renders low resolution data only.

It is worthwhile mentioning that view dependent rendering does not necessarily compromise quality. Due to the finite resolution of any display it is possible to achieve optimal rendering from image data at an appropriate resolution level whenever the display wouldn't reveal the highest resolution image details anyway. A view dependent renderer that exploits this can achieve very good large data rendering performance.

---

[1] Imaris typically uses blocks of 1MB size.

The blockwise multi-resolution format also facilitates the creation of performant analysis software for large images. BigStitcher [Hörl 2019] and ImarisStitcher are two applications for stitching of large amounts of image data that achieve high performance in part by making use of the blockwise multi-resolution data storage. Both applications compute the pairwise alignment between two images using phase correlation on lower resolution data to get robust alignment estimates and to get a speedup of a factor of 4 or more compared to running the same algorithm on high resolution data. Both alignment algorithms process only the overlapping regions of two images and thus profit from loading only those regions from the blockwise storage. Finally when the stitching applications copy the data from each image tile into a stitched output image a block-by-block procedure is an efficient method. These two stitching applications are just one example of how processing performance can benefit from a blockwise multi-resolution format. There are many other applications that can benefit similarly.

## 2 Streaming And Large Data Capability

The ImarisWriter library facilitates writing of image data that exceeds a computer's RAM by "streaming" the data to the library in small blocks. The library accepts data block by block, one block at a time. It is possible for example to send data of a 3D image to the writer slice by slice. More generally the library accepts 5-dimensional blocks of arbitrary size leaving the choice of input block size to the user. As shown in Figure 2 the library copies the data into internal memory blocks and pushes these through a multi-resolution resampling and compression pipeline to the output file "as soon as possible" to ensure minimal memory requirements. The library holds in RAM all internal blocks that are partially filled and pushes them to file as soon as they are full.

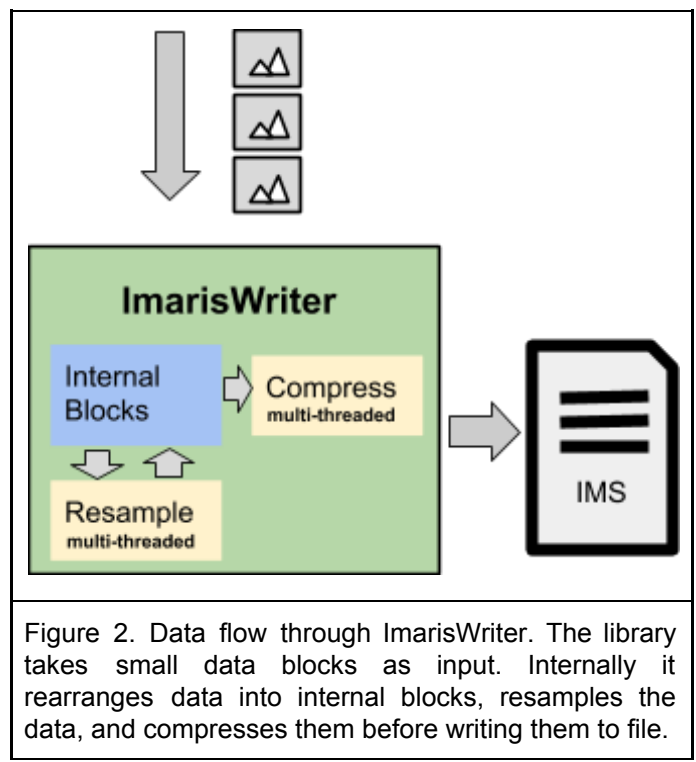

Figure 2. Data flow through ImarisWriter. The library takes small data blocks as input. Internally it rearranges data into internal blocks, resamples the data, and compresses them before writing them to file.

The user chooses the sequence in which blocks are streamed to the library by specifying the position of a block when it is copied. This "random access interface" has the effect that the maximum memory requirements of the library depend on the sequence in which the user streams data to it because this determines when internal blocks are full. That said, we demonstrate that peak memory requirements are often very low by listing some typical examples in Table 1.

| Image Size XYZCT | Total Input Memory | Input BlockSize | Input Sequence | Peak Memory Requirement |
|---|---|---|---|---|
| 2048x2048x512x3x1 | 12 GB | 512x512x1x1x1 | XYZCT | 373 MB |
| 2048x2048x512x3x1 | 12 GB | 512x512x1x1x1 | XYCZT | 1055 MB |
| 4096x4096x1024x3x1 | 96 GB | 512x512x1x1x1 | XYZCT | 1395 MB |
| 4096x4096x1024x3x1 | 96 GB | 512x512x1x1x1 | XYCZT | 4113 MB |
| 2048x2048x512x3x100 | 1200 GB | 512x512x1x1x1 | XYZCT | 373 MB |
| 2048x2048x512x3x100 | 1200 GB | 512x512x1x1x1 | XYCZT | 1055 MB |

Table 1. Measured Peak Memory Usage of ImarisWriter for various use cases.

The internal data blocks of the ImarisWriter library are three-dimensional in XYZ. Thus, to use the library in such a way that peak memory requirements are low, all that is necessary is to continuously fill XYZ space.

## 3 Performance

The ImarisWriter library is capable of writing compressed blockwise multi-resolution images with high speed. In just over 1 minute the library can write a 100 GB image to disk. To achieve this the library implements parallel compression and multi-resolution resampling. Figure 3 shows how parallelization improves the throughput of the writer. Making full use of all processors improves throughput from a poor single threaded 30 MB/s to 720 MB/s for Gzip compressed images. Much faster throughput can be achieved using LZ4 compression [LZ4 2011] and byte shuffling[2] with which the library achieves 1450 MB/s. This significantly outperforms uncompressed writing.

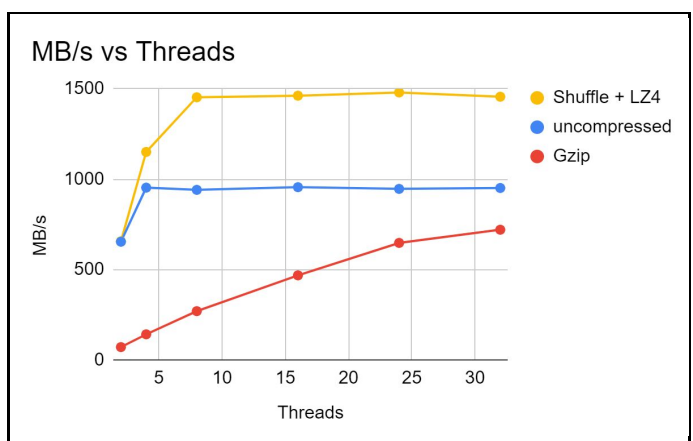

Figure 3. Performance of the ImarisWriter library as a function of the number of threads for different supported compression methods. Note that MB refers to the amount of data streamed to the library. The library adds a multi-resolution pyramid to these. The input image had XYZCT size (2048,2048,100,3,10) and a 16 bit data type amounting to 24000 MB of input data. The measurements reported here were made on a system with a 16 core AMD processor and an NVMe SSD under Microsoft Windows10.

---

[2] Starting with version 9.6 of the Imaris software the IMS format has the option to use LZ4-compression as well as byte shuffling followed by LZ4 compression. Both LZ4 compression and Shuffling are compatible with HDF5 filters. By Byte Shuffling or "Shuffle" we mean the reorganization of 16bit data to store consecutively the low precision bytes of multiple 16bit values and separately consecutively the corresponding high precision bytes.

The performance of our ImarisWriter library with Shuffle+LZ4 compression may well be sufficient to store data coming directly from a microscope on a typical storage device similar to the one used in our measurements. We believe that the Shuffle+LZ4 graph in Figure 3 suggests the possibility for higher throughput if the library is used with a faster storage device. In any case the performance of our library is significantly better than the performance of the KLB library [Amat 2015].

Side note on HDF5: The HDF5 library [Folk 2011] does not natively support parallel compression (or decompression). It does however provide API functions that allow programmers to build their own parallel compression and decompression on top of the HDF5 library[3]. Building such parallel compression functionality on top of the HDF5 library is not trivial and significantly overcomplicates the task of file writing. Our ImarisWriter library aims to provide a layer on top of HDF5 which takes care of all the challenges of high performance writing and is quite simple to use.

## 4 File Size

ImarisWriter provides the option to write images uncompressed or using one of several compression methods to reduce file size. Gzip compression with its nine different compression levels is a feature of the IMS format since its beginning in 2005. The possibility to compress using LZ4 compression is a new feature of the format. Both Gzip and LZ4 compression may be preceded by byte shuffling prior to compression. Table 2 provides a comparison of compression ratios for the different methods.

| Method | Compression Ratio | File Size [MB] |
|---|---|---|
| uncompressed | 1 | 33390 |
| Gzip level 2 | 2.36 | 14139 |
| Shuffle + Gzip level 2 | 2.92 | 11428 |
| LZ4 | 1.46 | 22767 |
| Shuffle + LZ4 | 2.11 | 15801 |

Table 2. Compression Ratio for different compression options. Compression ratios were measured on simulated data of 16 bit data type.

---

[3] H5DOwrite_chunk in version 1.8.11 and H5DOread_chunk in version 1.10.12 of HDF5

The absolute numbers in Table 2 cannot be representative of your microscopy data but it is our experience that the relative compression ratios of the different methods are quite stable for different image data, meaning for example that Shuffle + LZ4 is often significantly better than just LZ4.

## 5 Source Code and Licensing

The Source Code of ImarisWriter is available for download at the repository [ImarisWriter 2020]. The code is licensed under the Apache 2.0 license. The code is written in C++ and builds as a library with a C++ and a C API. The code is "small" and should be easily usable on many systems.

In addition to ImarisWriter itself we also provide code for a command line test program ImarisWriterTest at [ImarisWriterTest 2020]. This test program was used to do the performance measurements of section 4.

## 6 Usage

The ImarisWriter library can be used as follows. During its creation, the writer must be informed about the size of the image, the data type, and the size of the data blocks that will be copied to the library. Thereafter image data blocks are sent to the writer one by one (for example one could send data to the writer slice by slice). Finally when all image data have been sent, the writer can receive descriptive metadata before it is closed and file writing finishes.

The ImarisWriter library has a C++ API and a C-API. Using the C++-API pseudocode for writing a file mainly consists of a loop to copy all blocks to the library as shown in Listing 1.

```
bpImageConverter<bpUInt16> vImageConverter(...);

for ( all blocks ) {
            vImageConverter::CopyBlock(vBlockData, vBlockPosition);
}

vImageConverter::Finish(vParameters, ...);
```

Listing 1. C++ Pseudocode to write an image using the ImarisWriter library.

Even fully functional C++-code is short enough to print here in Listing 2.

```cpp
#include "imariswriter/interface/bpConverterTypes.h"
#include "imariswriter/interface/bpImageConverter.h"
#include <iostream>

using namespace bpConverterTypes;

void RecordProgress(bpFloat aProgress, bpFloat aTP) {std::cout << "Progress: " << aProgress*100 << "%\n";}

int main(int argc, char* argv[])
{
  tSize5D vImageSize(X, 2048, Y, 2048, Z, 100, C, 3, T, 1);
  tDimensionSequence5D vBlockDimensionSequence(X, Y, Z, C, T);
  tSize5D vBlockSize5D(X, 512, Y, 512, Z, 1, C, 1, T, 1);
  tSize5D vSample(X, 1, Y, 1, Z, 1, C, 1, T, 1);
  bpSize vBlockSize = vBlockSize5D[X] * vBlockSize5D[Y] * vBlockSize5D[Z];
  bpString vOutputFile = "C:\\ImarisWriterTest\\ImarisWriterTest.ims";
  cOptions vOptions;
  vOptions.mNumberOfThreads = 12;
  vOptions.mCompressionAlgorithmType = eCompressionAlgorithmGzipLevel2;

  bpUInt16* vFileBlock = new bpUInt16[vBlockSize];
  for (bpSize vVoxelIndex = 0; vVoxelIndex < vBlockSize; ++vVoxelIndex) {
    vFileBlock[vVoxelIndex] = vVoxelIndex % 512;
  }
```

```
  bpImageConverter<bpUInt16> vImageConverter(bpUInt16Type, vImageSize, vSample,
    vBlockDimensionSequence, vBlockSize5D, vOutputFile, vOptions, "ImarisWriterTest", "1.0", RecordProgress);

  for (bpSize vIndexT = 0; vIndexT < (vImageSize[T] - 1) / vBlockSize5D[T] + 1; vIndexT++) {
   for (bpSize vIndexC = 0; vIndexC < (vImageSize[C] - 1) / vBlockSize5D[C] + 1; vIndexC++) {
    for (bpSize vIndexZ = 0; vIndexZ < (vImageSize[Z] - 1) / vBlockSize5D[Z] + 1; vIndexZ++) {
     for (bpSize vIndexY = 0; vIndexY < (vImageSize[Y] - 1) / vBlockSize5D[Y] + 1; vIndexY++) {
      for (bpSize vIndexX = 0; vIndexX < (vImageSize[X] - 1) / vBlockSize5D[X] + 1; vIndexX++) {
       vImageConverter.CopyBlock(vFileBlock, {{X,vIndexX}, {X,vIndexY}, {Z,vIndexZ}, {C,vIndexC}, {T, vIndexT}});
      }
     }
    }
   }
  }

  tColorInfoVector vColorInfoPerChannel(vImageSize[C]);
  tTimeInfoVector vTimeInfoPerTimePoint;
  tParameters vParameters; vParameters["Image"]["ImageSizeInMB"] = "2400";
  vImageConverter.Finish({ 0,0,0,10,10,10 }, vParameters, vTimeInfoPerTimePoint, vColorInfoPerChannel, false);
  delete vFileBlock;
  return 0;
}
```

Listing 2. Fully functional C-code to write an image using the ImarisWriter library

## 7 Library Parameters

The following paragraphs explain some of the concepts used in the code of Listing 2.

### 7.1 Block Dimension Sequence

The two parameters named vBlockSize5D and vBlockDimensionSequence in the above example specify the storage layout of a data block that gets passed to the library in calls to CopyBlock. The dimension sequence is the five-dimensional generalization of what is sometimes called row-major or column-major in two dimensions. In fact, if block size XYZCT has values (512,512, 1,1,1), the choices (X,Y,Z,C,T) and (Y,X,Z,C,T) correspond to row-major and column-major layouts.

### 7.2 Block Position Index and Border Blocks

The position argument for image data blocks consists of five integers specifying the block index along every dimension. The first pixel/voxel of the block with position index (0,0,0,0,0) goes into the first pixel of the image, i.e. blocks are aligned to the beginning of the image. Border blocks at the end of the image extend outside the image when the image size is not exactly a multiple of block size along a dimension. The border blocks are required to contain padding data to fill the entire block size as declared.

### 7.3 Image Extent and Voxel Size

Instead of specifying voxel sizes in ImarisWriter one specifies ImageExtent with the min values in X, Y, and Z being the position (typically in micrometers) of the beginning (not the middle) of the first pixel/voxel and the position of the end (not the middle) of the last pixel/voxel along the respective dimension. Evidently voxel size may then be calculated from ImageExtent by (ExtentMaxX - ExtentMinX)/ImageSizeX.

### 7.4 Compression Options

The options argument specifies the compression method and the number of threads used for compression. The default compression method is Gzip level2 which is backward compatible with older versions of Imaris.

### 7.5 Option mForceFileBlockSizeZ1

The option mForceFileBlockSizeZ1 should be set to false almost always. It is intended only for those exceptional cases where the XY dimension of an image is extremely large to the point that it is not

6possible to keep much more than one layer entirely in RAM. In such cases this option may be used to reduce the memory requirements of ImarisWriter. Setting mForceFileBlockSizeZ1 has a negative impact on the performance of rendering and visualization and should thus be avoided.

### 7.6 Parameters / Metadata

The parameters passed to ImarisWriter in the Finish call are organized into sections, with each section being identified by a section name. In turn, each section contains different parameters organized in pairs, with a name and a value for each parameter. The purpose of this interface is to allow the addition of metadata to the output file before completion (e.g. a section named "AcquisitionDevice" and within that a parameter named "Model" with a value "DragonFly").

### 8 Adding a new Format to the library

It is relatively easy to add a new format to our library as long as this new format is compatible with the blockwise multi-resolution layout that our library produces[4]. The library internally has a file writer factory that creates writers derived from a filewriter interface. The main functionality of a filewriter is to write compressed data blocks to file one at a time. The library takes care of input data handling, chunking, multi-resolution resampling, and compression in a performant way and then dispatches file writing to one of the writer implementations. Formats that are compatible with blockwise multi-resolution handling can be added to our the library by implementing just the code that takes care of file writing. These writers naturally get all the performance benefits of the library's multi-threaded resampling and compression.

### 9 Summary

The open source ImarisWriter library is a high performance file writer for microscopy images. It creates image files suitable for high performance visualization and analysis. The library takes care of all the details of multi-resolution resampling, chunking, compression, multi-threading, etc and delivers its functionality to the user in a simple to use way.

---

[4] To a certain degree it is conceivable to add flexibility to the blockwise multi-resolution layout.

**References**

[Amat 2015] Amat, Fernando, Burkhard Höckendorf, Yinan Wan, William C. Lemon, Katie McDole, and Philipp J. Keller. "Efficient processing and analysis of large-scale light-sheet microscopy data." *Nature protocols* 10, no. 11 (2015): 1679.

[Dougherty 2009] Dougherty, Matthew T., Michael J. Folk, Erez Zadok, Herbert J. Bernstein, Frances C. Bernstein, Kevin W. Eliceiri, Werner Benger, and Christoph Best. "Unifying biological image formats with HDF5." *Communications of the ACM* 52, no. 10 (2009): 42-47.

[Folk 2011] Folk, Mike, Gerd Heber, Quincey Koziol, Elena Pourmal, and Dana Robinson. "An overview of the HDF5 technology suite and its applications." In *Proceedings of the EDBT/ICDT 2011 Workshop on Array Databases*, pp. 36-47. 2011.

[Goldberg 2005] Goldberg, Ilya G., Chris Allan, Jean-Marie Burel, Doug Creager, Andrea Falconi, Harry Hochheiser, Josiah Johnston, Jeff Mellen, Peter K. Sorger, and Jason R. Swedlow. "The Open Microscopy Environment (OME) Data Model and XML file: open tools for informatics and quantitative analysis in biological imaging." *Genome biology* 6, no. 5 (2005): R47.

[Hörl 2019] Hörl, David, Fabio Rojas Rusak, Friedrich Preusser, Paul Tillberg, Nadine Randel, Raghav K. Chhetri, Albert Cardona et al. "BigStitcher: reconstructing high-resolution image datasets of cleared and expanded samples." *Nature methods* 16, no. 9 (2019): 870-874.

[Imaris5 Format] Imaris5 File Format
https://github.com/imaris/ImarisWriter/blob/master/doc/Imaris5FileFormat.pdf

[ImarisWriter 2020] ImarisWriter Source Code Repository, https://github.com/imaris/ImarisWriter

[ImarisWritertest 2020] ImarisWriterTest Code, https://github.com/imaris/ImarisWriterTest

[LZ4 2011] LZ4 - Extremely fast compression, https://lz4.github.io/lz4/

[OME 6.0.0] Open Microscopy Format, [https://docs.openmicroscopy.org/ome-model/6.0.0/ome-tiff/specification.html](https://docs.openmicroscopy.org/ome-model/6.0.0/ome-tiff/specification.html)

[Pietzsch 2015] Pietzsch, Tobias, Stephan Saalfeld, Stephan Preibisch, and Pavel Tomancak. "BigDataViewer: visualization and processing for large image data sets." *Nature methods* 12, no. 6 (2015): 481-483.